\documentclass[amsmath,amssymb,aps,prb,twocolumn]{revtex4-2} 

\usepackage[usenames,dvipsnames]{xcolor}
\usepackage{graphicx,textcase} 
\usepackage{dcolumn,overpic} 
\usepackage{bm,color}
\usepackage[T1]{fontenc}
\usepackage{ae,aecompl}
\usepackage{chngcntr,gensymb}
\usepackage[para]{threeparttable}
\usepackage{etoolbox}
\usepackage{lipsum}
\usepackage[bottom]{footmisc}
\usepackage{setspace}

\AtEndEnvironment{table*}{\vskip10pt}{}{}
\usepackage{txfonts}
\usepackage{float}
\usepackage[normalem]{ulem}

\usepackage{url}
\usepackage{hyperref}
\hypersetup{colorlinks=true,breaklinks,linkcolor=blue,urlcolor=blue,citecolor=blue}

\begin{document}

\title{The impact of defects on excitations in two-dimensional bipartite uniaxial antiferromagnet insulators}

\author{Mahroo Shiranzaei}
\email{mahroo.shiranzaei@physics.uu.se}
 \affiliation{Division of Materials Theory, Department of Physics and Astronomy, Uppsala University, Box 516, SE-75120 Uppsala, Sweden}
\author{Jonas Fransson}
\email{jonas.fransson@physics.uu.se}
 \affiliation{Division of Materials Theory, Department of Physics and Astronomy, Uppsala University, Box 516, SE-75120 Uppsala, Sweden}

\begin{abstract}
We address scatterings of spin waves off uncorrelated defects in two-dimensional ($2$D) easy-axis antiferromagnet (AFM) insulators. Although an onsite magnetic anisotropy leads to gapped Goldstone modes, such that a long-range (AFM) order can be established in $2$D, lattice imperfections tend to weaken, and eventually destroy the magnetic ordering. Here, the impact of defects is considered within two limits, single and multiple defects. Using Green’s function, we perform self-consistent simulations to study magnon properties such as density of states and lifetime of induced resonances. Our findings show that repulsive defects decrease the magnon density of states while attractive ones may enhance it. We provide a comprehensive analysis of how defects can result in a reduction and even closing, the anisotropy-induced gap, which weakens the long-range (AFM) order parameter in the $2$D state. We conclude that a small concentration of random defects can fill the gap of magnon spectrum.
\end{abstract}

\maketitle

\section{\label{Intro}Introduction}
Magnetic fluctuations, like spin precession, are described by spin waves which can be regarded as dynamical magnetic eigenmodes. A quanta of spin waves, commonly known as magnons, are classified as low-energy Bosonic quasiparticles. Magnons carry linear and spin angular momentum as well as energy. Owing to the unique properties of magnons, an entirely new field has recently emerged, where the physics of magnons and their potential applications are investigated, aiming towards the design of magnon-based devices for information technology \cite{kruglyak2010magnonics, qin2015long}. One of the important advantages of magnons is their long coherence length, or lifetime, which opens the possibility for the transfer of data across distances much larger than what traditional electronic transfer technology is capable of \cite{kajiwara2010transmission, ross2019propagation, lebrun2018tunable}. For instance, data transfers carried by magnons have been achieved over distances ranging from tens of micrometers \cite{lebrun2018tunable} to several centimeters in the magnetic insulator Y$_3$Fe$_5$O$_{12}$ \cite{doi:10.1063/1.2834714}.

Antiparallel alignment of spins on two adjacent sublattices in antiferromagnets (AFMs) generates a vanishing zero net macroscopic magnetization. It makes AFMs robust against external magnetic perturbations which protect data from being changed \cite{RevModPhys.90.015005}. Besides, this feature gives the freedom to observe some effects like magnons with opposite circular polarizations useful to encode binary information \cite{cheng2016antiferromagnetic}, which are absent in ferromagnets (FMs) \cite{PhysRevLett.119.177202}. Moreover, the intrinsically fast magnetization dynamics in AFMs allow for operation at terahertz frequencies at room temperature \cite{PhysRevB.103.195144, qiu2021ultrafast, khymyn2017antiferromagnetic, kampfrath2011coherent}, something which is desirable for ultrafast applications \cite{wadley2016electrical, PhysRevLett.117.197201, stepanov2018long, qiu2021ultrafast} and quantum information \cite{PhysRevB.102.224418, PhysRevB.104.224302}.

Low-energy excitations of purely isotropic AFMs are governed by two degenerate modes with opposite helicities \cite{PhysRev.85.329, doi:10.1063/1.5109132}, in contrast to magnons in FMs in which magnon chirality is only right-handed with reference to the FM magnetization. This extra aspect enables chiral logic computing \cite{jia2021chiral}, magnon-mediated spin Hall \cite{PhysRevLett.117.217202, PhysRevLett.117.217203, PhysRevB.96.134425} and antiferromagnetic magnon field-effect transistor \cite{cheng2016antiferromagnetic}.

At nonzero temperature and owing to magnons, magnetic order is prevented in low dimensions; one ($1$D) and two dimensions ($2$D).
In 1966, Hohenberg \cite{ PhysRev.158.383}, and Mermin and Wagner \cite{ PhysRevLett.17.1133}, argued that the low-energy cost fluctuations of spins around their expectation values destroy magnetic order in $1(2)$D systems which are described by the isotropic Heisenberg Hamiltonian. As a consequence, constructing 2D magnets becomes challenging unless the spin fluctuations at the bottom of energy spectrum are suppressed. Moreover, dimensionality influences the magnon lifetimes which is a critical characteristic for magnon-based devices \cite{PhysRevLett.106.157204, PhysRevB.84.174418, qin2015long}.

The presence of an out-of-plane easy-axis magnetic anisotropy in magnetic materials can regenerate the long-range magnetic order in $1(2)$D \cite{lado2017origin}. However, the revived magnetic order is sensitive to lattice imperfections which makes the order unstable. Thus, a noteworthy debate is the study of lattice defects in low-dimensional structure by which translational invariance is broken. In particular, non-magnetic impurities can significantly alter the properties of low-dimensional magnets \cite{PhysRevB.94.054407,buczek2018spin, PhysRevB.94.174447,PhysRevB.104.024403}. Moreover, in addition to electron-magnon interactions, i.e., Landau damping \cite{PhysRevB.84.174418}, and spin-orbit coupling \cite{PhysRevLett.108.197205}, another mechanism that affects the magnon lifetime is scattering off impurities \cite{paischer2021eigenmodes}.

In Ref. \cite{izyumov1965some} it was shown that even at low concentrations of impurities in a ferromagnet, the emergence of local oscillations can give rise to a decrease in the magnetization if the induced energy level arises at the bottom of spin wave spectrum.
Furthermore, in a recent study of FMs on a honeycomb lattice \cite{PhysRevB.94.075401}, it was demonstrated that Dirac magnons arise and a local impurity induces a resonance in the magnon density of states, in a strong analogy with the effects of impurity scattering in graphene \cite{pan2000imaging, PhysRevLett.104.096804, PhysRevB.81.233405, PhysRevB.85.121103, PhysRevB.91.201411}. In addition, the position of the resonance approaches the Dirac point by increasing the impurity scattering potential.

In the current work, we investigate the impact of defects on the low-energy excitations in AFMs on a honeycomb lattices without and with a magnetic easy-axis anisotropy. We assume uncorrelated defects and consider them in two regimes, (i) single defect and (ii) multiple randomly distributed defects. By using the $T$-matrix technique, we develop the single particle dressed Green's function and calculate the local low-energy density of excitations. In the first case, a single defect, the $T$-matrix approach incorporates multiple magnon scattering processes off the single defect, thus, allowing for an exact solution. In the second case, multiple impurities, we employ spatial averaging and calculate the lowest-order self-energy diagrams within the self-consistent Born approximation, to evaluate the magnon lifetime and local density of excitations. This second component of the article is justified since it provides an averaged description of the physics, something which is relevant for optical and neutron scattering measurements, as well as any potential technology that might be exploited for AFMs.

We remark here, that we are not interested only in the magnon spectrum \emph{per se} but also in the distribution of the magnons between the lattices. It is well known, as we also briefly discuss in Sec. \ref{model}, that the bare magnon spectrum for AFMs is doubly degenerate, constituting two modes with opposite chirality. However, as one of the main targets in this article is to consider bipartite lattices with broken sublattice symmetry, it is preferable to investigate the physics in a sublattice-resolved framework. As we shall see, we gain additional information about the magnetic excitations in this form since it enables a perspective of magnons, not only as collective modes but also their composition in the lattice.

In summary, magnetic order in an interacting spin system can be destroyed by magnonic collective excitations of the structure. It essentially means that magnons quantize all possible spin deviations located in the lattice sites.
Our results show that each magnon mode distributes non-homogeneously between the two sublattices in AFMs, which is in stark contrast to the homogeneous partitioning of the modes between the sublattices in FMs. Moreover, it is seen that in the regime of a single defect with strong enough attractive potential, magnon local density of states increases compared to the MLDOS of a pristine lattice. By contrast, the repulsive defect diminishes the local density of magnons which indicates that a repulsive defect can suppress magnon effects in AFMs. In the case of easy-axis AFMs, an attractive single defect can induce a new peak within the energy gap of the system. These findings determine a reasonable range of impurity potentials by which the magnon gap can be filled in the regime of randomly distributed defects. It can be concluded that the existence of a diluted density of uncorrelated defects in $2$D easy-axis AFMs can prevent the establishment of a long-range order in AFMs due to easy-axis magnetic anisotropy.

The organization of the paper is as follows: Section \ref{model} reviews the model Hamiltonian of typical isotropic and easy-axis AFM insulators on a honeycomb lattice which is a non-Bravais bipartite lattice. We obtain the lowest-order magnon Hamiltonian by Holstein-Primakoff transformation. In section \ref{impurityeffect}, we study the effects of defects in introduced systems analytically by evaluating the single particle Green's function for the whole system including host and defect. Here, defects are considered within two regimes, a single defect, and randomly distributed defects. Thereafter, in Sec. \ref{results} we present our numerical findings of magnon density of states in the presence of defects. Finally, we conclude our results in Sec.~\ref{conclusion}. Furthermore, in Apps. \ref{AppA}--\ref{AppD} more details of calculations are provided.

\section{Modelling the anti-ferromagnetic structure}
\label{model}
\subsection{Hamiltonian}
\label{Hamiltonian}
An easy-axis AFM can be modeled by,
\begin{align}\label{Hspin}
H=& 
J \sum_{\langle i, j \rangle} \mathbf{S}_i \cdot \mathbf{S}_j
- \mathcal{K}_z \sum_{l \in \mathcal{A}, \mathcal{B}}( S^z_l)^2,
\end{align}
where $\mathbf{S}_{i(j)}$ denotes the spin vector at site $i(j)$ such that $\langle \mathbf{S}_{i \in \mathcal{A}} \rangle = (0,0,S)$ and $\langle \mathbf{S}_{j \in \mathcal{B}} \rangle = (0,0,-S)$; $\mathcal{A}, \mathcal{B}$, refer to different sublattices in a bipartite lattice. Here, $J$ is the isotropic exchange coupling between the nearest neighbors, indicated by $\langle i,j \rangle$, and assumed to be anti-ferromagnetic, $ J > 0 $. Moreover, $ \mathcal{K}_z > 0$ denotes the easy-axis on-site anisotropy, forcing the spins to be collinearly aligned perpendicular to the material surface.

In order to model Bosonic excitations in this structure, we apply the Holstein-Primakoff (HP) transformation \cite{PhysRev.55.1218}, which for the AFM is given by \cite{PhysRev.87.568, robert2007quantum, auerbach2012interacting},
\begin{subequations}
\label{HP}
\begin{align}
&S^z_i = S - a^\dag_i a_i ,
&S^-_i = a^\dag_i \sqrt{2S - a^\dag_i a_i}, 
&&S^+_i = \big( S^-_i \big)^\dag ,
\\
&S^z_j = - S + b^\dag_j b_j ,
&S^-_j = \sqrt{2S - b^\dag_j b_j}b_j,
&&S^+_j = \big( S^-_j \big)^\dag ,
\end{align}
\end{subequations}
in which $a^\dag_i (a_i)$ is the creation (annihilation) operator where $i \in \mathcal{A}$, and $b^\dag_j (b_j)$ refers to the one where $j \in \mathcal{B}$. The HP transformation for AFMs stresses that $a$ and $b$ carry opposite spin angular momentum along $\hat{z}$-axis on different sublattices. At low enough temperature, the number of excitations is very small, $\langle a^\dag_i a_i \rangle, \langle b^\dag_i b_i \rangle \ll 2S$, \textit{i.e.}, the spin fluctuations are small around the classical direction, $|S^z| \simeq |S|$.
Using the HP transformation makes it possible to rewrite the spin Hamiltonian, Eq.~\eqref{Hspin}, in terms of Bosonic operators describing spin deviations around the equilibrium state.
By expanding the square root in Eq.~\eqref{HP} in powers of $1/S$, we construct a quadratic model $H = E^{0} + H^{(2)}_b$. The classical energy of the ground state is given by $E^0 = -N S^2 (z J/2 + \mathcal{K}_z)$ at zero temperature, where $N$ is the total number of sites in the lattice. For a lattice with periodic boundary conditions, we introduce the Fourier transformation as,
\begin{subequations}
\label{FT}
\begin{align}
a_i =& \sqrt{\frac{2}{N}} \sum_q e^{i \, \mathbf{q} \cdot \mathbf{r}_i} a_{\mathbf{q}} \; ,
\hspace{0.2cm}
i \in \mathcal{A},
\\
b_j =& \sqrt{\frac{2}{N}} \sum_q e^{i \, \mathbf{q} \cdot \mathbf{r}_j} b_{\mathbf{q}} \; ,
\hspace{0.2cm}
j \in \mathcal{B},
\end{align}
\end{subequations}
thereafter the quadratic Hamiltonian in reciprocal space is obtained,
\begin{align}\label{nonH}
H^{(2)}_b
    =&
    \sum_q
    \Big[
    \varepsilon_0 a^\dag_{\mathbf{q}} a_{\mathbf{q}}
    +
    \varepsilon_0 b^\dag_{\mathbf{q}} b_{\mathbf{q}}
    \nonumber\\&
    +
    S Z J \gamma_{-\mathbf{q}} a_{\mathbf{q}} b_{-\mathbf{q}}
    +
    S Z J \gamma_{\mathbf{q}}
    b^\dag_{-\mathbf{q}}a^\dag_{\mathbf{q}} 
    \Big],
\end{align}
where $\varepsilon_0 = S (Z J + 2 \mathcal{K}_z)$. Here, $\gamma_{\mathbf{q}} =  \sum_{i=1}^Z e^{i \mathbf{q} \cdot \boldsymbol{\delta}_i} /Z$ is the lattice structure factor where $Z$ denotes the number of nearest neighbours and $\boldsymbol{\delta}_i$ is the distance vectors to the nearest neighbours. The structure factor $\gamma_{\bf q}$ for a 2D square lattice ($Z=4$) is $\gamma_{\bf q}=2[\cos aq_x+\cos aq_y]/z$ and for a bipartite hexagonal ($Z=3$) lattice $\gamma_{\bf q}=e^{iaq_x}[1+2 e^{-i3aq_x/2}\cos\sqrt{3}aq_y/2]/z$, where $a$ is the lattice constant.
In the spinor basis $\Psi^\dag_\mathbf{q} = (a^\dag_\mathbf{q}, b_{-\mathbf{q}}) $, we rewrite the Hamiltonian as $ H^{(2)}_b = \sum_\mathbf{q} \Psi^\dag_\mathbf{q} h_\mathbf{q} \Psi_\mathbf{q} $ where,
\begin{align}
h_\mathbf{q}
=
\begin{pmatrix} 
    \varepsilon_0 & S Z J \gamma_\mathbf{q} \\
    S Z J \gamma_{-\mathbf{q}} & \varepsilon_0
\end{pmatrix}.
\end{align}

Conventionally, the magnon spectrum is considered by diagonalizing $H_b^{(2)}$ using the Bogoliubov transformation \cite{kittel1963quantum, nolting2009quantum}; in appendix \ref{AppA} we consider our model in the Bogoliubov transformed coordinates for the sake of clarity. However, an alternative is provided by retaining the reference to the sublattices and considering the equation of motion for the excitation Green's function \cite{RevModPhys.44.406, PhysRevB.25.432}. This approach gives the opportunity to obtain the spectrum of excitations and the projection of spin deviations at each sublattice \cite{izyumov1965impurity}.
Moreover, as we investigate the magnon density of states in both the bare and perturbed AFMs, as well as extract the contributions of different sublattices and the effect of defects, it is convenient to remain in the lattice coordinates and employ the Green's function formalism. The Green's function for the sublattice resolved magnons is defined by,
\begin{align}
{\bf G}(\varepsilon, \mathbf{q}, \mathbf{q}')
=
- i \langle\langle \Psi_\mathbf{q} | \Psi^\dag_{\mathbf{q}'} \rangle\rangle
=
- i
\left\langle\left\langle
\begin{pmatrix}
a_\mathbf{q} \\
b^\dag_{-\mathbf{q}}
\end{pmatrix}
\Bigg|
\begin{pmatrix}
a^\dag_{\mathbf{q}'} \ b_{-\mathbf{q}'}
\end{pmatrix}
\right\rangle\right\rangle.
\end{align}
where $\langle\langle ...\rangle\rangle$ denotes the magnon propagator in energy space. The equation of motion in energy space reads as
$
\varepsilon \langle\langle A B \rangle\rangle
=
\langle\langle [A,B] \rangle\rangle
+
\langle\langle [A,H] B \rangle\rangle
$ where $H$ is the system Hamiltonian and $A, B$ are arbitrary operators \cite{nolting2008fundamentals}. Note that although Green's function is not diagonal in a sublattice basis, the commutator on the right-hand side of equation of motion tells us that the coupling between excitations on different sublattices is taken into account. Performing the equation of motion on the bare Green's function results in,
\begin{align}\label{AFMgreen}
{\bf G}_0(\varepsilon,{\bf q})=&
\begin{pmatrix}
\varepsilon - \varepsilon_0 & -S Z J \gamma_\mathbf{q}
\\
S Z J \gamma_{-\mathbf{q}} & \varepsilon + \varepsilon_0
\end{pmatrix}^{-1}
\begin{pmatrix}
1 & 0
\\
0 & -1
\end{pmatrix}
\nonumber\\&
=
\frac{1}{\varepsilon^2-\varepsilon_0^2+|S Z J \gamma_{\bf q}|^2}
\begin{pmatrix}
\varepsilon + \varepsilon_0 & -S Z J \gamma_\mathbf{q}
\\
-S Z J \gamma_{-\mathbf{q}} & -\varepsilon + \varepsilon_0
\end{pmatrix}
	,
\end{align}
from this expression, it is vividly clear that the (diagonal) components representing the ${\cal A}$- and ${\cal B}$-sublattices are not equivalent; $\bf G_0^\mathcal{B B}$ is evaluated from $\bf G_0^\mathcal{A A}$ by $\varepsilon \rightarrow -\varepsilon$.

It is, however, important to remember that the Green's function $\bf{G}_0^{\cal{BB}}$ does not represent magnons in the $\cal{B}$-sublattice, but rather the anti-excitations in the same, c.f., electron-hole pairs. This means that while the operator $a^\dagger_{\bf q}$ creates an excitation in the $\cal A$-sublattice, the operator $b^\dagger_{-\bf q}$ annihilates an excitation in the $\cal B$-sublattice. It is, furthermore, important to recall that the magnons, referred to as $\alpha$- and $\beta$-magnons are constructed as superpositions of the operators $a_{\bf q}$ and $b^\dagger_{-\bf q}$, and $b_{\bf q}$ and $a^\dagger_{-\bf q}$, respectively (See App. ~\eqref{AppA}).

The consequences of choosing the operator basis $\Psi_{\bf q}$ can be seen by writing the Green's function on the form of,
\begin{subequations}
\begin{align}
{\bf G}_0(\varepsilon,{\bf q})=&
	\frac{1}{\varepsilon-\varepsilon_{\bf q}^\alpha}
	\begin{pmatrix}
		|u_{\bf q}|^2 & -S Z J \gamma_\mathbf{q} \\
		-S Z J \gamma_\mathbf{-q} & -|v_{\bf q}|^2
	\end{pmatrix}
\label{eq-G1}
\\&
	+
	\frac{1}{\varepsilon+\varepsilon_{\bf q}^\alpha}
	\begin{pmatrix}
		|v_{\bf q}|^2 & S Z J \gamma_\mathbf{q} \\
		S Z J \gamma_\mathbf{-q} & -|u_{\bf q}|^2
	\end{pmatrix}
\label{eq-G2}
	,
\end{align}
\end{subequations}
where $|u_{\bf q}|^2=(1+\varepsilon_0/\varepsilon_{\bf q})/2$ and  $|v_{\bf q}|^2=(1-\varepsilon_0/\varepsilon_{\bf q})/2$, such that $|u_{\bf q}|^2-|v_{\bf q}|^2=1$, whereas the energy $\varepsilon_{\bf q}^\alpha$ is given in Eq. \eqref{baredisp} below. The partitioning of the Green's function into two terms tells us that the magnon, which is a positive energy excitation, is retained from the first, Eq. \eqref{eq-G1}, whereas the second term, Eq. \eqref{eq-G2}, accounts for the excitations at energies $\varepsilon<0$. Therefore, the bare Green's function $g_\alpha$ for the $\alpha$-magnons, associated with this basis, is given by the trace of Eq. \eqref{eq-G1}, that is,
\begin{align}
g_\alpha(\varepsilon,{\bf q})=&
	\frac{|u_{\bf q}|^2-|v_{\bf q}|^2}{\varepsilon-\varepsilon_{\bf q}^\alpha}
	=
	\frac{1}{\varepsilon-\varepsilon_{\bf q}^\alpha}
	.
\end{align}
However, this also shows that the upper left (lower right) of Eq. \eqref{eq-G1} accounts for the portion of the magnon which is confined to the ${\cal A}\ ({\cal B})$-sublattice and is done so with the weight $|u_{\bf q}|^2$ ($-|v_{\bf q}|^2$). The $\beta$-magnon has an equivalent representation in which the roles of the $\cal A$- and $\cal B$-sublattices are interchanged. Retaining this partitioning of the description enables a more detailed perspective of the impurity scattering, as we shall see below.

\begin{figure}[ht]
\begin{center}
    {\includegraphics[width=0.49\columnwidth]{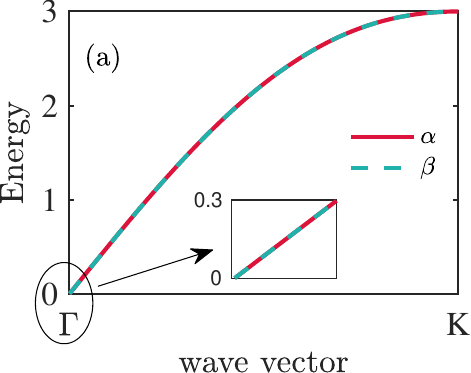}} 
    {\includegraphics[width=0.49\columnwidth]{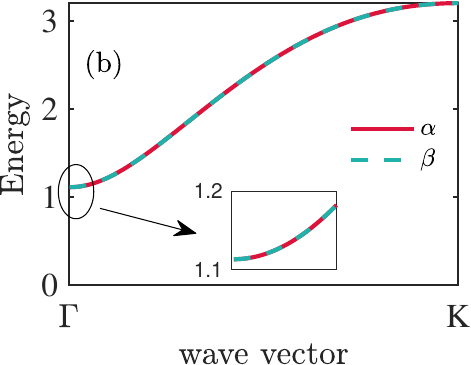}} 
\end{center}
\caption{(Color online) Energy spectrum for (a) an isotropic AFM and (b) an easy-axis AFM with $\mathcal{K}_z = 0.1 J$. The used parameters are $J = 1$, $S=1$. The inset in panel (a) and (b) shows the linear and quadratic behaviour of dispersion for isotropic and easy-axis AFMs respectively.}
    \label{fig:dispersion}
\end{figure}

The dispersion of the free magnons is obtained from the zeros of the denominator of ${\bf G}_0$ as
\begin{equation}\label{baredisp}
\varepsilon_\mathbf{q}^{\alpha/\beta} =
\sqrt{
\varepsilon_0^2 
- 
S^2 Z^2 J^2 |\gamma_\mathbf{q}|
^2
}
.
\end{equation}
which are degenerate for the two magnon branches ($\varepsilon^{\alpha}_\mathbf{q}$, $ \varepsilon^{\beta}_{\mathbf{q}}$) but can be distinguished by their opposite chirality as they conduct opposite spin angular momentum \cite{PhysRevLett.117.217202}. Thereafter, the free magnon Hamiltonian can be written as,
\begin{equation} \label{Hdiag}
\mathcal{H}^{(2)}
=
E_0^q
+
\sum_\mathbf{q}
\varepsilon^{\alpha}_{\mathbf{q}} \,
\alpha^\dag_\mathbf{q} \alpha_\mathbf{q}
+
\varepsilon^{\beta}_\mathbf{q} \,
\beta^\dag_\mathbf{q} \beta_\mathbf{q},
\end{equation}
where $\alpha_{\bf q}$ and $\beta_{\bf q}$ are magnon eigenmodes which are collective modes of spin fluctuations on sublattice $\mathcal{A}$ and $\mathcal{B}$ in the collinear AFM. Additionally, the quantum correction to the ground-state energy is $E_0^q = - N  \varepsilon_0 /2$. In summary, magnons in the AFM are circularly polarized even in the presence of an easy-axis anisotropy.  In the absence of external fields, an easy-axis anisotropy does not break the $U(1)$ symmetry, which leads to that magnons distribute over two degenerate bands with opposite helicities \cite{PhysRev.85.329, doi:10.1119/1.1933416, doi:10.1063/1.5109132}.

From Eq.~\eqref{baredisp}, it is clear that for an isotropic AFM ($\mathcal{K}_z = 0$), the energy dispersion is linear, $ \varepsilon_{\mathbf{q}} = \zeta q /\sqrt{2}$, where $\zeta = S Z\,J\,a$, as shown in Fig.~\ref{fig:dispersion} (a), where the magnon energy of the isotropic AFM is plotted as a function of $q$ near $\Gamma$. In this case, magnons are massless with group velocity depending on the exchange interaction as $ v_g = \zeta/\sqrt{2}$. These aspects are universal fingerprints of Dirac-type materials \cite{doi:10.1080/00018732.2014.927109}. 
Note that the gapless band structure of magnons sustains the emergence of Goldstone excitations, corresponding to spin deviations around the equilibrium, which can occur without any cost of energy. Hence, at finite temperatures, such easily excited fluctuations destroy long-range order in $2$D isotropic AFMs.

In both the square and honeycomb lattices, the structure factor $|\gamma_\mathbf{q}|\approx 1- (a q)^2/4$, for small $\mathbf{q}$ around the center of Brillouin zone, i.e., $\Gamma$-point. It is clear from Eq.~\eqref{baredisp} that a finite easy-axis anisotropy, $\mathcal{K}_z>0$, induces a gap at the bottom of the magnon spectrum around the $\Gamma$-point. The gap opening is accompanied by the introduction of a non-linearity of the energy dispersion, which is shown in Fig.~\ref{fig:dispersion} (b). The gap opening stabilizes the system and revive long-range magnetic order in the 2D structure. In this case, the AFM resonance frequency, evaluated at $\mathbf{q} =0 $, is given by $\omega_\text{RAFM} =  4 ZJ  \mathcal{K}_z + 4 \mathcal{K}_z^2  $. The dispersion relation in Eq.~\eqref{baredisp}, shows that despite $\varepsilon_\mathbf{q}^{\alpha/\beta}$ depends on the lattice structure factor $\gamma_{\bf q}$, which is unique for the lattice, in general,  the magnon dispersion is qualitatively similar in square and honeycomb lattices for ${\bf q}\rightarrow0$. For small $\mathbf{q}$ and sufficiently weak anisotropy, $\mathcal{K}_z \ll J$, the magnon bands are defined through the relations,
\begin{subequations}
\begin{align}
\varepsilon^{\alpha/\beta}_\mathbf{q}
\approx&
 2 S \sqrt{Z J \mathcal{K}_z} \biggl(1 + \frac{Z J a^2}{16 \mathcal{K}_z} q^2\biggr).
 \end{align}
\end{subequations}
Here, due to the anisotropy, the non-zero effective mass and group velocity of magnons are, respectively,
\begin{subequations}
\begin{align}
\big|m^*\big| =& \hbar^2 \biggl(\frac{\partial^2 \varepsilon_q}{\partial q^2}\biggr)^{-1}
    = \frac{4 \hbar^2}{\zeta a}\sqrt{\frac{\mathcal{K}_z}{Z J}},
\\
\big|v_g\big| =& \frac{\partial \varepsilon_q}{\partial q}
    = \frac{\zeta a}{4}\sqrt{\frac{Z J}{\mathcal{K}_z}} \,q. 
\end{align}
\end{subequations}

\subsection{Bare magnon local density of states}
\label{GreenF}

We finish this section with a further analysis of Eq. \eqref{AFMgreen}, from which we obtain the magnon local density of states for bare systems in both isotropic and easy-axis AFMs.

The density of magnon states $n$ is related to the Green's function by the relation $n(\varepsilon, \mathbf{q}) = -{\rm Im}{\rm Tr}\, {\bf G}(\varepsilon, \mathbf{q})/\pi$, where the trace runs over sublattice degrees of freedom. The bipartite structure allows us to write $n(\varepsilon, \mathbf{q}) = \sum_x n^x(\varepsilon, \mathbf{q})$, where $n^x(\varepsilon, \mathbf{q})= -{\rm Im} \, G^{xx}/\pi$, $x={\cal A},\ {\cal B}$, indicating the sublattice-resolved density of magnon states which enables us to study the two sublattices contribute to the total magnon DOS.

Magnons are eigenmodes of the spin-wave spectrum and should be regarded as a collective behaviour in a spin structure. In this sense, they involve all the spin deviations in the lattice. However, here we are interested in how the magnon LDOS is distributed over the two sublattices. Hence, to address the LDOS of the bare structures, we consider the on-site Green's function given by $g_0(\varepsilon) \sigma_0= \sum_\mathbf{q} {\bf G}_0(\varepsilon,\mathbf{q}) / N $ \cite{PhysRevB.73.125411} where $\sigma_0$ denotes $2\times2$ Identity matrix. In the following, we express magnon LDOS for unperturbed isotropic and easy-axis AFMs.

\subsubsection{Isotropic AFM}
For AFMs without anisotropy, the $g_0^{\mathcal{A A}}=g_0^{\mathcal{A A}}(\varepsilon)$ component of  on-site Green's function reduces to,
\begin{align}\label{g0AA}
    g_0^{\mathcal{A A}} =&
    \frac{\varepsilon + ZJ S}{W^2}
    \Biggl(
    \ln\biggl|\frac{\varepsilon^2}{\varepsilon^2 - \zeta^2 q_c^2/2}\biggr|
    -
    i\pi\text{sgn}(\varepsilon)\Theta\Bigl(\zeta^2 q_c^2/2 - \varepsilon^2\Bigr)
    \Biggr),
\end{align}
where $\text{sgn}(\varepsilon)$ refers to the sign function, and $q_c$ denotes the cut-off wave vector which relates to the magnon bandwidth, $W$, through the relation of $ W = v_g q_c $. Moreover, $\Omega = 1/{\cal S}$, where ${\cal S} = 3 \sqrt{3} a^2/2$, is the area of the hexagonal unit cell, such that we can set $ W^2= 4 \pi \Omega v^2_g $ \cite{PhysRevB.73.125411}. The integrated magnon densities of states in the ${\cal A}$- and ${\cal B}$-sublattice, respectively, are obtained from the imaginary part of the sum over all k-points in Brilloun zone of Eq.~\eqref{AFMgreen}, hence, giving
\begin{subequations}\label{bareAFM}
\begin{align}
    n_0^{\mathcal{A}}(\varepsilon)
    =&
    \frac{\varepsilon + ZJS}{W^2}
    \text{sgn}(\varepsilon) \, \Theta\Bigl(\zeta^2 q_c^2/2 - \varepsilon^2\Bigr)
    ,
\\
n_0^{\mathcal{B}}(\varepsilon)
    =&
    \frac{-\varepsilon + Z\,J\, S}{W^2}
    \text{sgn}(\varepsilon) \, \Theta\Bigl(\zeta^2 q_c^2/2 - \varepsilon^2\Bigr)
    .
\end{align}
\end{subequations}
These expressions imply that total density of magnon states, $n_0(\varepsilon)$ vanishes at the zero energy point since $sgn(\varepsilon=0) = 0$. In addition, it shows $n_0^{\mathcal{A}}$ and $n_0^{\mathcal{B}}$ increases and decreases, respectively, by increasing energy.
Indeed, a comparison of the imaginary parts of $g_0^{\cal AA}$ and $g_0^{\cal BB}$  which are plotted in Fig. \ref{fig:g0} (a), along with their corresponding real parts, as a function of the energy indicates that the contribution of two sublattices are completely different in $n_0(\varepsilon)$ which is in stark contrast to FM honeycomb lattices \cite{PhysRevB.94.075401}.

\subsubsection{Easy-axis AFM}
\begin{figure}[ht]
\begin{center}
{\includegraphics[width=0.95\columnwidth]{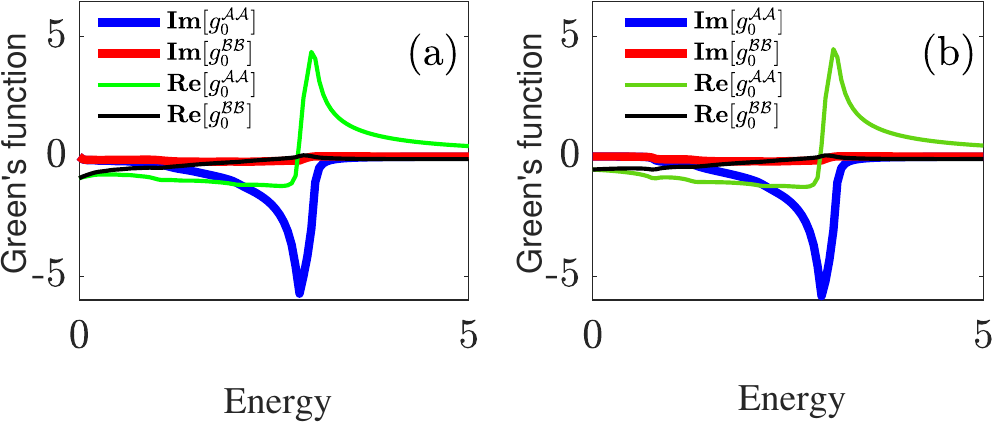}}
\end{center}
\caption{Real and imaginary part of on-site Green's function, $g_0^{\mathcal{AA}}$ and $g_0^{\mathcal{BB}}$, where we set in (a) $\mathcal{K}_z = 0$  and in (b) $\mathcal{K}_z = 0.05 J$. The other parameters are $J = 1$, $S=1$. The solid and dotted lines illustrate the imaginary and real part of on-site Green's function.}
    \label{fig:g0}
\end{figure}

A non-vanishing easy-axis anisotropy, ${\cal K}_z>0$, opens up a low energy gap in the magnonic structure. This property can be seen from the Green's functions of sublattice $\mathcal{A}$ and $\mathcal{B}$, which are given by the expression
\begin{subequations}\label{easyGF}
\begin{align}
g_0^{\mathcal{A}\mathcal{A}}
=&
- \frac{\varepsilon + \varepsilon_0}{W^2}
\Bigg[ 
\ln{\bigg|1- \frac{\zeta^2 q^2_c/2}{\varepsilon^2 -\tilde{\cal K}}\bigg|}
\nonumber\\&
+
i \, \pi \, 
\text{sgn} (\varepsilon) \,
\Theta\Big(\varepsilon^2 - \tilde{\cal K}\Big) \,
\Theta\Big(\zeta^2 q^2_c/2-\varepsilon^2+\tilde{\cal K}\Big)
\Bigg],
\\
g_0^{\mathcal{B}\mathcal{B}}
=&
\frac{\varepsilon - \varepsilon_0}{W^2}
\Bigg[ 
\ln{\bigg|1- \frac{\zeta^2 q^2_c/2}{\varepsilon^2 -\tilde{\cal K}}\bigg|}
\nonumber\\&
+
i \, \pi \, 
\text{sgn} (\varepsilon) \,
\Theta\Big(\varepsilon^2 - \tilde{\cal K}\Big) \,
\Theta\Big(\zeta^2 q^2_c/2-\varepsilon^2+\tilde{\cal K}\Big)
\Bigg],
\end{align}
\end{subequations}
where $\tilde{\cal K}=4 Z J \mathcal{K}_z S^2$ and $\varepsilon_0 = S (Z J + 2 \mathcal{K}_z)$. In Fig.~\eqref{fig:g0}(b), the real and imaginary parts of the onsite Green's function are plotted as a function of energy. Compared to the isotropic AFM, Fig.\eqref{fig:g0}(a), the properties of the magnonic structure are here shifted to positive energies, reflecting the gap introduced by the anisotropy.

\section{Impurity effects}
\label{impurityeffect}
The existence of defects in materials can induce either localized modes or resonances in the energy spectrum. Specifically, a localized mode has a small, if any, overlap with the host band structure while a resonance resides inside the continuous part of the spectrum. Such resonances can, in magnon systems, be detected in experiments by optical measurements and neutron scattering \cite{RevModPhys.44.406}. 
The impurities can be considered in different regimes which are, here, theoretically investigated by using the single-particle Green's function. In this section, we first consider a single defect and study how the magnon density of states is affected. Then, we proceed with multiple randomly distributed impurities. In this case, we explore the lifetime of localized states generated by impurities. We finalize this section with a comprehensive analysis of the impact of defects on elementary excitations in AFMs.

\subsection{Single defect}
\label{singledefect}
By manipulating one spin in one of the sublattices, such local perturbation can be considered as a local defect, which may influence the rest of the system. For sake of concreteness, we consider a single defect located at the coordinate ${\bf r}_0$ in sublattice $\mathcal{A}$, modeled by $H_{d} = U_0(\sigma_0+\sigma_z)a^\dag_0a_0/2$. Here, $\sigma_\alpha$, $\alpha\in\{0,x,y,z\}$, define the $2\times2$ Identity and Pauli matrices in pseudospin space.
In the presence of the local defect, the Green's function can be partitioned as
\begin{align}\label{Gd}
    {\bf G}(\varepsilon,\mathbf{q},\mathbf{q}^\prime) = 
    {\bf G}_0(\varepsilon,\mathbf{q},\mathbf{q}^\prime)
    +
    \delta{\bf G}(\varepsilon,\mathbf{q},\mathbf{q}^\prime),
\end{align}
where ${\bf G}_0(\varepsilon,\mathbf{q},\mathbf{q}^\prime) $ is the bare Green's function, whereas $\delta{\bf G}(\varepsilon,{\bf q},{\bf q}')$ is the correction due to the local perturbation. The dressed Green's function manifests a propagation of excitations through the perfect system in which scattering occurs by a local perturbation. Here, we calculate this correction using the $T$-matrix expansion,
\begin{align}
        \delta{\bf G}(\varepsilon,\mathbf{q},\mathbf{q}^\prime)
        =
        {\bf G}_0(\varepsilon,\mathbf{q}){\bf T}(\varepsilon){\bf G}_0(\varepsilon,\mathbf{q}^\prime),
\end{align}
where the T-matrix is given by,
\begin{align}
    {\bf T}(\varepsilon)
    =&
    \Bigl(
        H^{-1}_d - g_0(\varepsilon)
    \Bigr)^{-1}
    =
    H_d
    \Bigl(
        \sigma_0 -g_0(\varepsilon)H_d
    \Bigr)^{-1},
\end{align}
where $g_0(\varepsilon)$ is the on-site Green's function \cite{PhysRevB.73.125411}.
After straightforward calculations, the diagonal components of the correction term, $\delta{\bf G}(\varepsilon,\mathbf{q})\equiv\delta{\bf G}(\varepsilon,\mathbf{q},{\bf q})$, are obtained,
\begin{subequations}\label{deltaG}
\begin{align}
        \delta \bf G^{\mathcal{A}\mathcal{A}}(\varepsilon,\mathbf{q})
        =&
        \frac{1}{U^{-1}_0-g^{\mathcal{AA}}_0}
        \frac{(\varepsilon + \varepsilon_0)^2} 
        {\Big(
        \varepsilon^2
        - \varepsilon^2_0 + S^2 Z^2 J^2 |\gamma_\mathbf{q}|
^2
        \Big)^2},
\\
        \delta \bf G^{\mathcal{B}\mathcal{B}}(\varepsilon,\mathbf{q})
        =&
        \frac{1}{U^{-1}_0-g^{\mathcal{AA}}_0}
        \frac{S^2 Z^2 J^2 |\gamma_\mathbf{q}|^2} 
        {\Big(
        \varepsilon^2
        - \varepsilon^2_0 + S^2 Z^2 J^2 |\gamma_\mathbf{q}|^2
        \Big)^2}
        ,
\end{align}
\end{subequations}
providing the corrections to the density magnon of states,
\begin{subequations}\label{deltaN}
\begin{align}
        \delta n^{\mathcal{A}} (\varepsilon,\mathbf{q})
        =&
        \frac{-1}{\pi}
        \frac{{\rm Im}~g^{\mathcal{AA}}_0}{|U^{-1}_0-g^{\mathcal{AA}}_0|^2}
        \frac{(\varepsilon + \varepsilon_0)^2}{\Big(
        \varepsilon^2
        - \varepsilon^2_0 + S^2 Z^2 J^2 |\gamma_\mathbf{q}|^2
        \Big)^2},
\\
        \delta n^{\mathcal{B}}(\varepsilon,\mathbf{q})
        =&
        \frac{-1}{\pi}
        \frac{{\rm Im}~g^{\mathcal{AA}}_0}{|U^{-1}_0-g^{\mathcal{AA}}_0|^2}
        \frac{S^2 Z^2 J^2 |\gamma_\mathbf{q}|^2}{\Big(
        \varepsilon^2
        - \varepsilon^2_0 + S^2 Z^2 J^2 |\gamma_\mathbf{q}|^2
        \Big)^2}.
\end{align}
\end{subequations}
From the above equations, we determine the energy of the impurity resonance by requiring
\begin{align}\label{inducedpeak}
|U^{-1}_0-g^{\mathcal{AA}}_0|^2 
	\Big(\varepsilon^2 - \varepsilon^2_0 + S^2 Z^2 J^2 |\gamma_\mathbf{q}|^2 \Big)^2 =&
	0
	.
\end{align}
This condition implies either that ${\rm Im}~g^{\mathcal{AA}}_0 = 0$ and ${\rm Re}~g^{\mathcal{AA}}_0 = 1/U_0$, simultaneously, or $\varepsilon^2 - \varepsilon^2_0 + S^2 Z^2 J^2 |\gamma_\mathbf{q}|^2 =0$.
The zeros arising from the former requirement correspond to impurity-induced resonances in the magnon density, Eq. \eqref{deltaN}. For these resonances to emerge in the density of the spin excitations, the defect has to be attractive, $U_0 < 0$ .
An attractive scattering potential enhances the spin deviations compared to the repulsive scattering potential, which leads to the enlarged production of magnons. Hence, repulsive scattering potentials stabilize the formation of a gapped magnon density and suppress the resonance peak at low energy. For more details see Sec. \eqref{repulsive}.

\subsection{Multiple defects}
\label{randomdefects}
Beyond the case of a single defect, as a more realistic scenario, we consider here the effect of multiple defects randomly distributed on the honeycomb lattice. Considering a collection of randomly distributed impurities is motivated since many experimental techniques, as well as potential technological applications, sample the averaged properties of the AFMs, rather than the very local effects that can be addressed for single defects. Hence, by assuming a dilute concentration of defects, correlations between them can be neglected, while at the same time the averaged properties of the AFM can be taken into accont. We investigate the problem of  short-ranged scattering potentials, and we assume an equal number, $n$, of independent defects in the two sublattices constituting the volume $V$; all with the same scattering strength $U$. Under this assumption, we model the defects through
\begin{subequations}
\begin{align}
    	c U \sigma_\mathcal{A}=&
	 \frac{n U}{V}\frac{\sigma_0+\sigma_z}{2}
,
\\
    	c U \sigma_\mathcal{B}=&
	\frac{n U}{V}\frac{\sigma_0-\sigma_z}{2}
,
\end{align}
\end{subequations}
on sublattice $\mathcal{A}$ and $\mathcal{B}$, respectively, where $c=n/V$ is the concentration of defects. In addition, the scattering potential is assumed to be ${\bf q}$-independent. Distributing defects in the lattice breaks the translational invariance in the system. However, since we assume uniform randomly distributed defects, we can make a spatial averaging over the defects, which has the property of recovering the translational invariance of the magnon Green's function \cite{haug2008quantum}, which is outlined in the following.

The spatial averaging of the collection of scattering defects $H_{{\bf qq}'}=cU\sum_{i\in{\cal A,B}}\sigma_ie^{-i({\bf q}-{\bf q}')\cdot{\bf r}_i}$ gives rise to,
\begin{align}
     \bar{H}_{\mathbf{q}\mathbf{q}^\prime}
     =&
     c U \,  \delta_{\mathbf{q}\mathbf{q}^\prime} \sigma_0.
 \end{align}
Then, the dressed Green's function can be written as the Dyson equation \cite{haug2008quantum},
 \begin{align}
     \bf G(\varepsilon,\mathbf{q},\mathbf{q}^\prime)
     =
     &
     \delta_{\mathbf{q}\mathbf{q}^\prime}
     {\bf G}_0(\varepsilon,\mathbf{q})
 \nonumber\\&
     +
     {\bf G}_0(\varepsilon,\mathbf{q})
     \frac{1}{\Omega} \sum_{\mathbf{k}i}
     H_{\mathbf{k}} e^{-i(\mathbf{p}-\mathbf{k})\cdot\mathbf{R}_i}
     {\bf G}(\varepsilon,\mathbf{k},\mathbf{q}^\prime).
 \end{align}
By retaining the first-order correction of the Green's function, we obtain,
 \begin{align}
     \bar{\bf G}_1(\varepsilon,\mathbf{q},\mathbf{q}^\prime)
     =
     \delta_{\mathbf{q}\mathbf{q}^\prime}
     cU
     {\bf G}^2_0(\varepsilon,\mathbf{q})
     .
 \end{align}
The second order correction $\bar{\bf G}_2$ provided through the averaging procedure, $ \overline{H_{\mathbf{q k}} g_{\mathbf{k}} H_{\mathbf{k}{\mathbf{q}}^\prime}}$, is given by \cite{haug2008quantum, somphonsane2020universal},
\begin{align}
     \bar{\bf G}_2(\varepsilon,\mathbf{q})
     =&
     \Bigl(
		c^2 U^2
		(1-1/N)
		\bf G_0(\varepsilon,\mathbf{q})
		+
		\Sigma(\varepsilon,\mathbf{q})\,
	\Bigr)
	\bf G_0^2(\varepsilon,\mathbf{q})
	,
 \end{align}
where the self energy within self-consistent Born approximation is defined by,
 \begin{align}
     \Sigma(\varepsilon)
     =&
     \,c\, U^2
     \sum_{i = \mathcal{A, B}}
     \sum_{\mathbf{q}} 
     \sigma_i
     \bar{\bf G}(\varepsilon,\mathbf{q})
     \sigma_i.
 \end{align}
Thereafter, the dressed Green's function is self-consistently evaluated by
$\bar{G}(\varepsilon, \mathbf{q}) =G_0(\varepsilon, \mathbf{q})-G_0(\varepsilon, \mathbf{q})\Sigma[\bar{G}]\bar{G}(\varepsilon,{\bf q})$. In general, the self energy can be written as $\Sigma(\varepsilon,\mathbf{q}) = \Lambda(\varepsilon,\mathbf{q}) - i/2\tau(\varepsilon,\mathbf{q})$. The real part shifts the position of the resonance, while the imaginary part provides a broadening of the resonant state which is proportional to the inverse of the resonant lifetime $\tau$. The summation over the sublattices yields the diagonal self-energy
 \begin{align}
     {\bf \Sigma}
     =&
     \,c\, U^2
     \sum_{\mathbf{q}}
     \begin{pmatrix}
     \bar{G}^{\mathcal{AA}}(\mathbf{q}) & 0
     \\
     0 & \bar{G}^{\mathcal{BB}}(\mathbf{q})
     \end{pmatrix}
=
     \begin{pmatrix}
     \Sigma^{\mathcal{AA}} & 0
     \\
     0 & \Sigma^{\mathcal{BB}}
     \end{pmatrix}.
\end{align}
 from which the lifetime is determined as $1/\tau = - 2\text{Im} \text{Tr}{\bf \Sigma}(\varepsilon)$.

\section{Numerical results}
\label{results}
Here, we proceed by analyzing the integrated magnon DOS, $n(\varepsilon)=n^{\cal A}(\varepsilon)+n^{\cal B}(\varepsilon)$, where $n^x(\varepsilon)=\sum_{\bf q}n^x(\varepsilon,{\bf q})/N$.  We present our numerical findings about effects of both single and multiple randomly distributed scattering potentials.

\subsection{Pristine lattice}
\label{Inducedresonancepeak}

\begin{figure}[ht]
\begin{center}
    {\includegraphics[width=0.49\columnwidth]{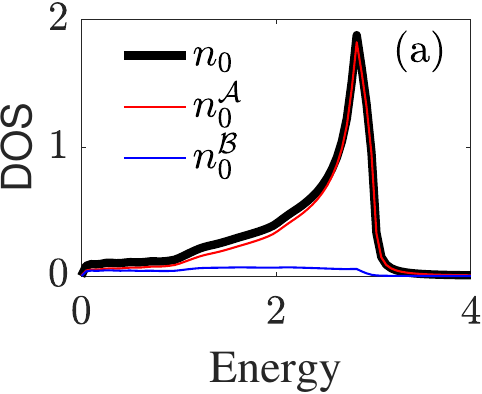}} 
    {\includegraphics[width=0.49\columnwidth]{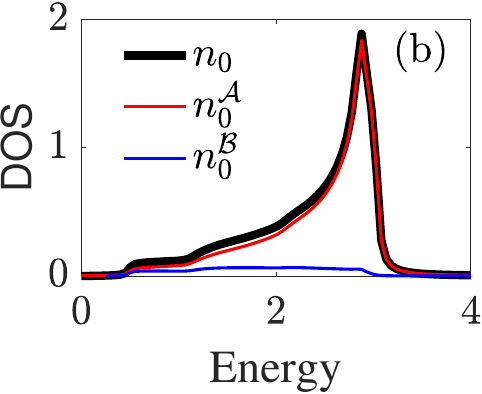}}
\end{center}
\caption{Integrated magnon DOS for the pristine AFM on a honeycomb lattice. Panels (a) and (b) display magnon DOS for the isotropic ($K_z = 0$) and uniaxial ($K_z = 0.02J$) AFMs corresponding to the imaginary part of Eqs. \eqref{g0AA} and \eqref{easyGF} respectively. In these panels, black line shows the total DOS while red and blue lines represent the contribution of sublattice $\mathcal{A}$ and $\mathcal{B}$ in total DOS. We use $J = 1$, $S=1$.}
\label{fig:AFMhoneycombBare}
\end{figure}

In Fig. \ref{fig:AFMhoneycombBare}, the integrated DOS for the $\alpha$-magnon is plotted for (a) pristine isotropic and (b) easy-axis AFMs. 
The signature of anisotropy in panel (b) is seen as a gap opening. The total DOS of the $\alpha$-magnon in the pristine lattice is not equally distributed between the two sublattices, except at $\varepsilon=0$, as shown in Fig.~\ref{fig:AFMhoneycombBare} (a), (b), where the sublattice-resolved DOS are represented by red and blue lines. For low energies, the contributions of both sublattices to the total DOS vary linearly. With increasing energy, however, the $\cal A$-sublattice density, $n_0^\mathcal{A}$, strongly increases whereas the $\cal B$-sublattice density, $n_0^\mathcal{B}$, remains nearly constant. The picture for the $\beta$-magnon is a mirror image of that of the $\alpha$-magnon.

The distinct attributes of the sublattice-resolved DOS stress that the bosonic excitations from different sublattices ($a$ and $b^\dagger$) have non-identical roles for the magnon eigenmodes ($\alpha$ and $\beta$). This effect happens because the excitations in sublattice $\mathcal{A}$ correspond to the spin in this sublattice gradually becomes less well-defined until it cannot be defined any longer in a meaningful sense. As a consequence of the AFM exchange interaction between the two sublattices, the vast majority of the $\alpha$-magnon remains within the ${\cal A}$-sublattice with only a minor leakage into the ${\cal B}$-sublattice, which is indicated by the small $n_0^\mathcal{B}$. 
This feature is in the Hamiltonian for the AFM represented by the coupling term between excitations on two sublattices of the form $a_{\bf q} b_{-\bf q}$ and $b^\dag_{-\bf q} a^\dag_{\bf q}$. These terms show that the creation of excitations in one of the sublattices does not entail an equivalent excitation in the other.

\begin{figure}[t]
\begin{center}
    {\includegraphics[width=0.48\columnwidth]{fig3a.pdf}} 
    {\includegraphics[width=0.49\columnwidth]{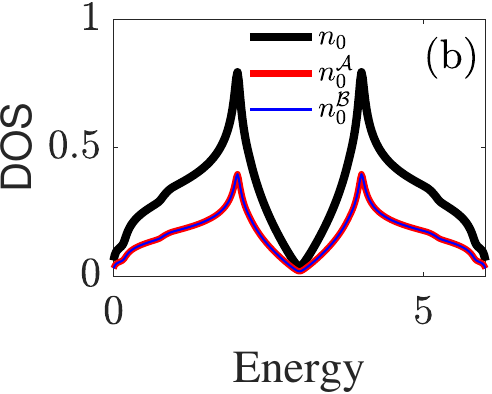}}
\end{center}
\caption{Integrated magnon DOS for the pristine isotropic AFM and FM on a honeycomb lattice are shown in panel (a) and (b) respectively. In these panels, black line shows the total magnon DOS while red and blue line present the contribution of sublattice $\mathcal{A}$ and $\mathcal{B}$ in total DOS. We use $J = 1$, $S=1$, and.}
\label{fig:FMhoneycombBare}
\end{figure}

By contrast, in a bipartite FM lattice (See the FM Hamiltonian in App. \eqref{AppC}) the bosonic excitations are equally distributed between the two sublattices. This is illustrated in Fig. \ref{fig:FMhoneycombBare}, where we show a comparison of DOS for an (a) AFM and (b) FM honeycomb lattice.
It comes from this fact that in bipartite FM, the coupling between $a_{\bf q}$ and $b_{\bf q}$ appears as $a^\dag_{\bf q} b_{\bf q}$ and $b^\dag_{\bf q} a_{\bf q}$ known as splitting terms. Splitting terms show that generating excitations on one sublattice leads to excitations on the other sublattice (See App. \eqref{AppC} for more details about FM on bipartite lattices). That is, the deviation of one spin propagates equally throughout the whole lattice, creating magnon eigenmodes under FM interactions.

\subsection{Induced resonance peak}
\label{Inducedresonancepeak}
 
\begin{figure}[t]
\begin{center}
    {\includegraphics[width=0.49\columnwidth]{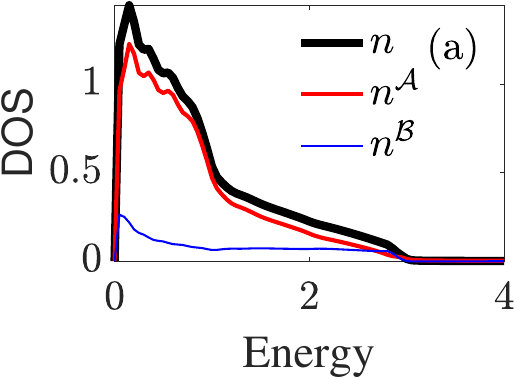}} 
    {\includegraphics[width=0.46\columnwidth]{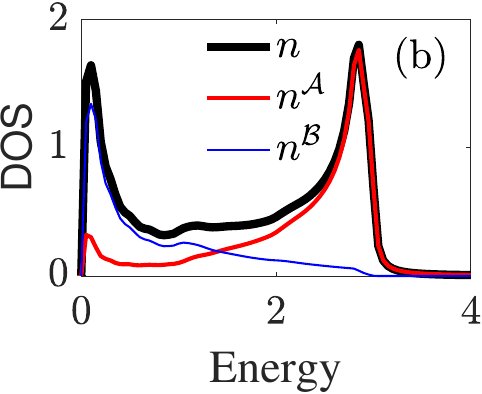}}
\end{center}
\caption{Integrated magnon DOS for the perturbed isotropic AFM with a single defect on a honeycomb lattice. Panels (a) and (b) display magnon DOS for the cases where the defect is located on sublattice $\mathcal{A}$ and $\mathcal{B}$ respectively. In these panels, solid black line show the total DOS while red and blue lines represent the contribution of sublattice $\mathcal{A}$ and $\mathcal{B}$ in total DOS. We use $J = 1$, $S=1$, and $U_0 = -1.3$.}
\label{fig:AFMhoneycomb}
\end{figure}

\begin{figure}[t]
\begin{center}
    {\includegraphics[width=0.49\columnwidth]{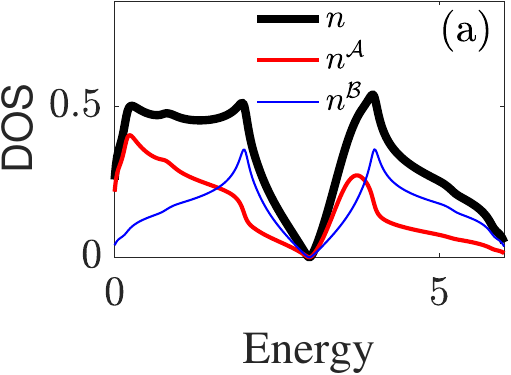}} 
    {\includegraphics[width=0.49\columnwidth]{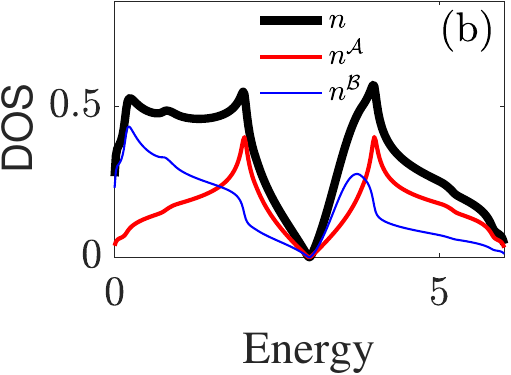}}
\end{center}
\caption{Integrated magnon DOS for the perturbed isotropic FM with a single defect on a honeycomb lattice. Panels (a) and (b) display magnon DOS for the cases where the defect is located on sublattice $\mathcal{A}$ and $\mathcal{B}$ respectively. In these panels, solid black line show the total DOS while red and blue lines represent the contribution of sublattice $\mathcal{A}$ and $\mathcal{B}$ in total DOS. We use $J = 1$, $S=1$, and $U_0 = -1$.}
\label{fig:FMhoneycomb}
\end{figure}

The presence of a single defect changes the magnon densities dramatically. The momentum-integrated magnon DOS for AFM and FM honeycomb lattices is plotted in Fig. \ref{fig:AFMhoneycomb} and Fig. \ref{fig:FMhoneycomb}, respectively. The plots in Fig. \ref{fig:AFMhoneycomb} (a)  and Fig. \ref{fig:FMhoneycomb} (a) correspond to configurations with a single defect located in sublattice $\mathcal{A}$ while Fig. \ref{fig:AFMhoneycomb} (b)  and Fig. \ref{fig:FMhoneycomb} (b) a single defect is located in sublattice $\mathcal{B}$. As can be concluded from a comparison with the bare magnon densities, see Figs. \ref{fig:AFMhoneycombBare}, \ref{fig:FMhoneycombBare}, adding a single defect leads to a substantial modification of the magnon density.

Considering the AFM, an attractive defect ($U<0$) introduces magnon resonances in both densities $n^{\cal A}$ and $n^{\cal B}$, see Fig.~\ref{fig:AFMhoneycomb}. The resulting total magnon density also strongly depends on which sublattice the single defect is located in. The density is hugely redistributed whenever the defect is located in the sublattice $\cal A$, Fig.~\ref{fig:AFMhoneycomb} (a), which is connected to that the large density at high energies in the pristine lattice is reshuffled by the introduction of the defect. Since the density in sublattice $\cal B$ is generally much smaller, the total density reshuffling caused by the defect becomes nominally much less should the defect be located in this sublattice, Fig.~\ref{fig:AFMhoneycomb} (b).

Summarizing the result for the AFM honeycomb lattice, it can be observed that the response of bosonic excitations to the perturbation in isotropic AFMs reveals that, for low energy, the defect on sublattice $\cal A$ gives rise to leakage of states into sublattice $\cal B$. On the other hand, locating a single defect in sublattice $\mathcal{B}$ induces a strong low energy resonance in $n^\mathcal{B}$. This behavior is traced back to the emergence of the strong impurity resonance in the ${\bf q}$-resolved LDOS of sublattice ${\cal B}$, see Eq. \eqref{deltaN}. This result is in contrast to what one observes in a honeycomb lattice, e.g., graphene and FM Dirac magnons \cite{PhysRevB.94.075401} and other Dirac materials \cite{doi:10.1080/00018732.2014.927109, PhysRevB.97.180402}.

By contrast, in FMs the overall properties of the total magnon DOS is independent of which sublattice the single defect is located in, as shown in Fig. \ref{fig:FMhoneycomb}. Here, the resulting magnon density for the configuration with a single defect in sublattice $\cal A$ indicates that state is redistributed around low energies in which the bare density $n_0^\mathcal{A}$ play the most important role, see Fig. \ref{fig:FMhoneycomb} (a). A relocation of the defect to sublattice $\cal B$ results in the same total density, however, with the modifications in the respective sublattice interchanged.

\subsection{Induced localized peak}
\label{Inducedlocalizedpeak}

\begin{figure}[t]
\begin{center}
    {\includegraphics[width=0.46\columnwidth]{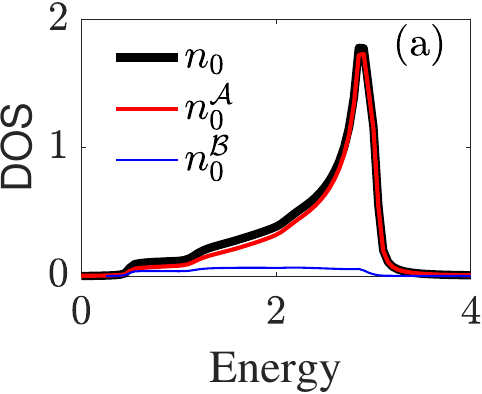}} 
    {\includegraphics[width=0.49\columnwidth]{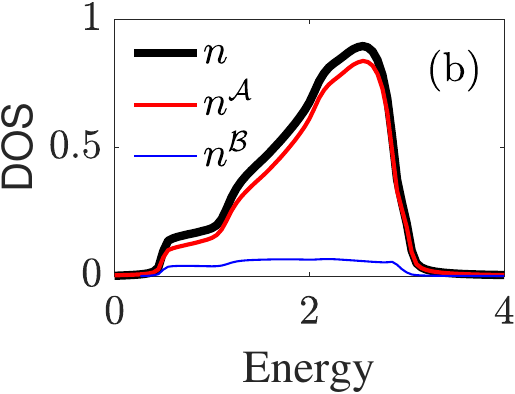}}
    {\includegraphics[width=0.49\columnwidth]{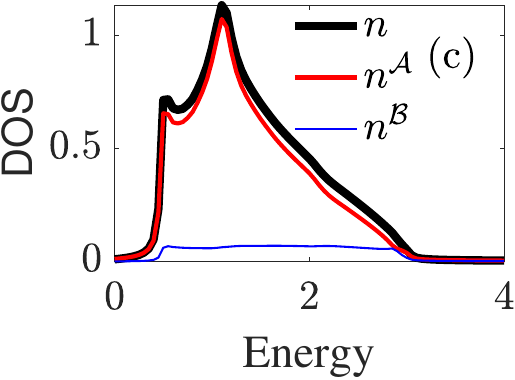}}
    {\includegraphics[width=0.49\columnwidth]{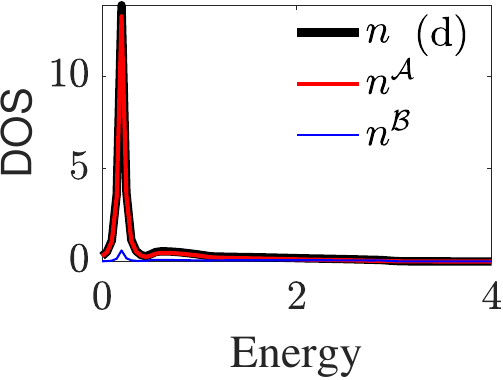}}
\end{center}
\caption{Integrated magnon DOS for the AFM on a honeycomb lattice. Panel (a) refers to the case of pristine lattice while panels (b, c and d) demonstrate the effect of a single defect in sublattice $\mathcal{A}$ with $U_0 = -0.3, -0.9, -1.5$ on magnon DOS respectively. In these panels, solid black line shows the total DOS. Red and blue lines represent the contribution of sublattice $\mathcal{A}$ and $\mathcal{B}$ in the total DOS. We use $J = 1$, $S=1$, and $K_z = 0.02J$.}
\label{fig:AFMhoneycombUs}
\end{figure}

In easy-axis AFMs, the magnon gap opens up due to the magnetic anisotropy, $\mathcal{K}_z>0$ which can be seen in Fig.~\ref{fig:AFMhoneycombUs} (a). Apart from the gapped magnon DOS, the same observations as discussed in the isotropic AFMs remain valid here. These observations are: in analogy with the isotropic case, the magnon DOS is not equally distributed between the sublattices; introduction of a single attractive defect in the lattice alters the resulting magnon DOS, see panels (b)--(d). 
The plots in Fig. \eqref{fig:AFMhoneycombUs}, furthermore, illustrate that an increasing scattering potential amplitude $U_0$ associated with the single defect shifts the magnon DOS to the edge of the gap. For sufficiently large potential $U_0$, a localized level emerges in the gap. The location of this localized level can be obtained from the correction term of density given in Eq. \eqref{deltaN}.
Futhermore in presence of the easy-axis anisotropy, the density of magnon states at the localized level becomes stronger in sublattice $\mathcal{A}$, where the defect is located, than in sublattice $\cal B$. This feature is in stark contrast to the result of Dirac FMs and more similar to the case of gapped Dirac materials like thin films of topological insulators \cite{PhysRevB.95.235429}.

\begin{figure}[t]
\begin{center}
    {\includegraphics[width=0.49\columnwidth]{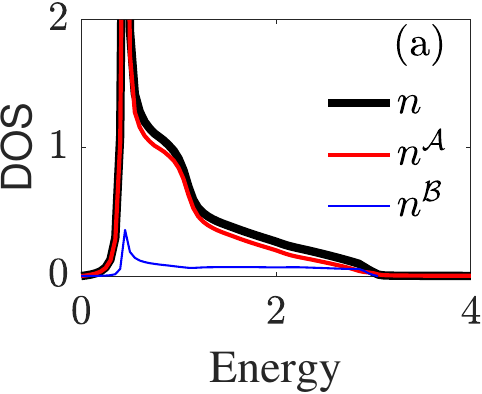}}
    {\includegraphics[width=0.49\columnwidth]{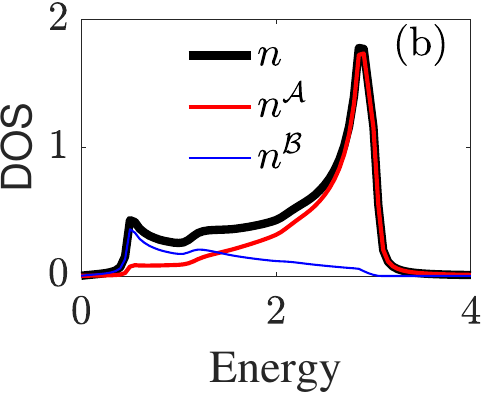}}
\end{center}
\caption{Integrated magnon DOS for the perturbed easy-axis AFM on a honeycomb lattice with a single defect. Panels (a) and (b) display magnon DOS for the case where a single defect located in sublattice $\mathcal{A}$ and $\mathcal{B}$, respectively. In these panels, solid black line shows the total DOS while red and blue dotted lines represent the contribution of sublattice $\mathcal{A}$ and $\mathcal{B}$ in total DOS. We use $J = 1$, $S=1$, $K_z = 0.02J$ and $U_0 = -1.2$.}
\label{fig:AFMhoneycombKz02AB}
\end{figure}

Like previously, it matters largely in which sublattice the defect is located. In Fig.~\ref{fig:AFMhoneycombKz02AB}, the integrated density of magnon states are plotted as a function of energy for a single defect in the (a) $\mathcal{A}$-sublattice and (b) $\mathcal{B}$ -sublattice, in an easy-axis AFM on a honeycomb structure. For a strong enough defect potential, a localized level emerges either at the edge of or inside the gap. Nevertheless, a single defect located in the $\mathcal{A}$-sublattice leads to a stronger density redistribution in the vicinity of the gap.

Scattering off a repulsive defect in the easy-axis AFM, tend to close the gap at the bottom of the spectrum. Consequently, the long-range order provided by the anisotropy, is destroyed by the impurity scattering. This can be regarded as the scattering leading to the reintroduction of the Goldstone modes which were gapped out by the anisotropy. However, in order to analyze this situation more in depth, we go beyond a configuration with a single defect and consider the effect of scattering off multiple defects.

 \subsubsection{repulsive scattering potentials}\label{repulsive}

\begin{figure}[b]
\begin{center}
    {\includegraphics[width=1\columnwidth]{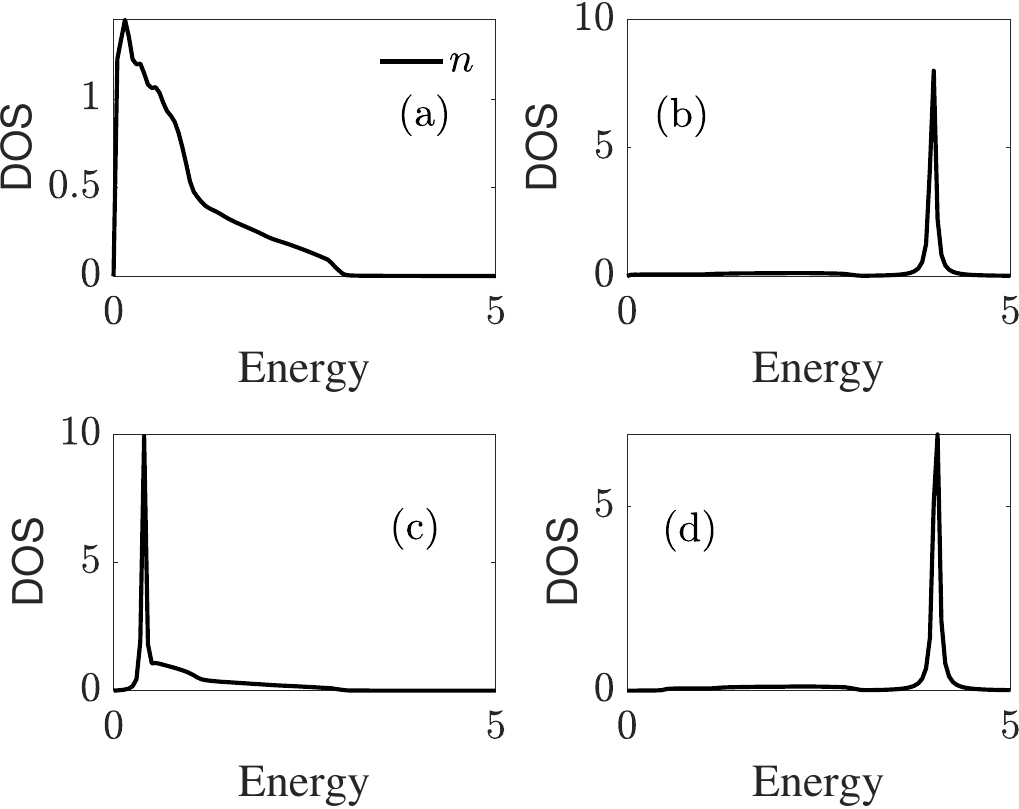}} 
\end{center}
\caption{Integrated magnon DOS for the perturbed AFM with a single defect on a honeycomb lattice. Panel (a,b) and (c,d) display integrated magnon DOS for the isotropic and the uniaxial AFMs with $K_z = 0.02J$. System includes a single defect with an attractive potential $U_0=-1.3$ in panel (a,c) while with a repulsive potential $U_0=+1.3$ in panel (b,d). In these panels, solid black lines show the total DOS. We use $J = 1$, $S=1$.}
\label{fig:attractiverepulsive}
\end{figure}

As can be seen in Fig. \eqref{fig:g0} that to satisfy the condition of ${\rm Im} \, g_0^\mathcal{AA} = 1/U_0$, in low energy, attractive scattering potential is a prerequisite. Fig. \eqref{fig:attractiverepulsive} shows that a positive scattering potential reduces magnon DOS compared to the bare system, which can be interpreted that the spin system becomes stable against magnons in low energy in the presence of a repulsive single defect. In panels (a,c), resonance and induced peaks are respectively induced due to an attractive single defect while panel (b,d) shows that the repulsive defect suppresses the magnon density of states in low energy.

\subsection{Random defects}
\label{multipledefects}

\begin{figure}[t]
\begin{center}
    {\includegraphics[width=0.49\columnwidth]{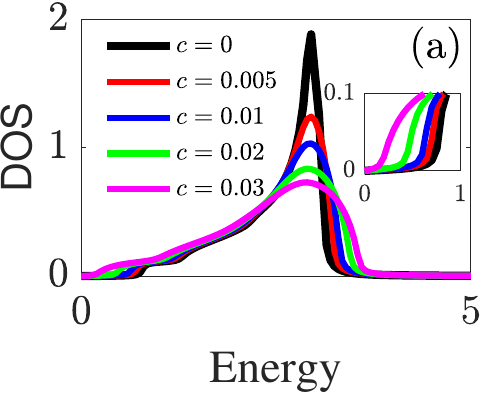}} 
    {\includegraphics[width=0.49\columnwidth]{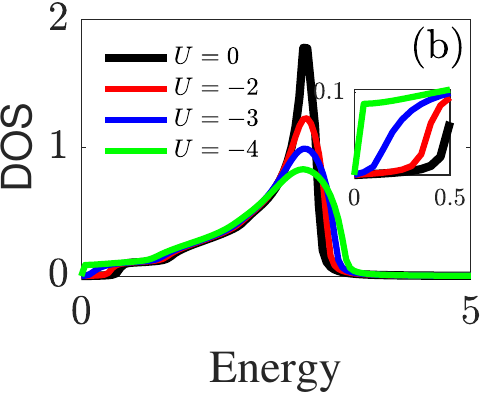}} 
\end{center}
\caption{Total magnon DOS in AFMs randomly diluted by defects. In panel (a), magnon DOS in terms of energy is illustrated for different concentrations of defects ($c = 0, 0.5, 1, 2, 3 \%$) where $\mathcal{K}_z = 0.05 J$ and defect strength is $U=-2$. Panel (b) shows the impact of the strength of defect on magnon DOS where $c=5 \%$.}
\label{fig:randomdos}
\end{figure}

In Fig.~\ref{fig:randomdos} (a), we plot the total magnon DOS for the easy-axis AFM including multiple defects as a function of energy for different defect concentrations. Each defect gives rise to a local enhancement of the magnon density. The averaged effect of each such enhancement leads to a full closing of the magnon gap for large enough defect concentrations. Fig.~\ref{fig:randomdos} (b), illustrates the impact of defect strength on the magnon gap Despite a larger anisotropy makes the density gap wider, our results suggest that the ordered state in the easy-axis AFMs is practically sensitive to the impurity scattering.

\begin{figure}[t]
\begin{center}
\includegraphics[width=0.49\columnwidth]{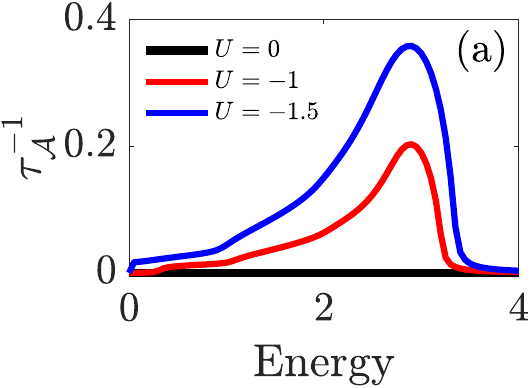} 
\includegraphics[width=0.49\columnwidth]{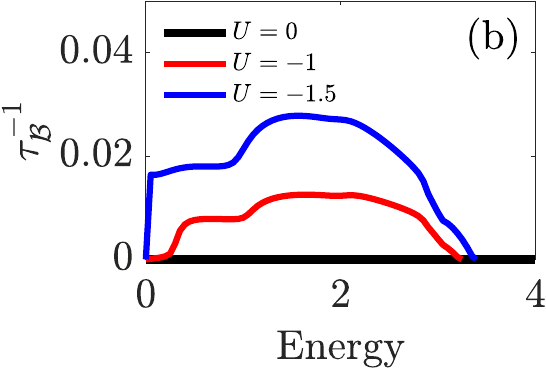}
\end{center}
\caption{Inverse of magnon lifetime on sublattice (a) $\mathcal{A}$ and (b) $\mathcal{B}$ are illustrated as a function of energy in a system randomly diluted by defects with concentration ($c = 3 \%$) and different strength ($U = 0, -1, -1.5$) while other parameters are as $J=1, K_z = 0.02$.}
\label{fig:lifetime}
\end{figure}

 The panels in Fig.~\ref{fig:lifetime} display the sublattice resolved lifetimes of the spin excitations as a function of energy for easy-axis AFMs on a honeycomb lattice. As ${\tau}^{-1} = \tau^{-1}_\mathcal{A} +  \tau^{-1}_\mathcal{B} $, in general, one can see that lifetime of magnons decreases by increasing energy. In lower energy, magnons have a longer lifetime. The enhancement of the defect strength reduces the overall relaxation time of magnons. Moreover, it can be seen that fluctuations in the sublattice ${\cal A}$ tend to have a shorter lifetime (larger $\tau^{-1}_{\mathcal{A}}$) than in sublattice ${\cal B}$. The averaged magnon lifetimes are measurable by, e.g., inelastic neutron scattering \cite{PhysRevB.3.157, PhysRevB.4.2280} and nuclear magnetic resonance \cite{PhysRevB.7.307}.

\section{Conclusions}
\label{conclusion}
In conclusion, here we report that the magnetic order in two-dimensional AFM lattice structure is very sensitive to magnonic scattering off defects. While the absent magnetic order in the isotropic AFM lattice can be restored through an easy-axis anisotropy, scattering off local attractive impurities tends to generate localized levels in the gap below the magnon band. By considering randomly distributed scattering defects throughout the lattice, we find that even small defect concentrations lead to the gap closing and, hence, destroy the AFM order. Moreover, in presence of scattering defects, the magnon excitations acquire an overall reduced relaxation time suggesting that also the stability of the magnons is reduced. This observation indicates, in turn, that for a viable information transfer using magnons in two-dimensional AFM, one should consider materials with very few defects and impurities to sustain reliable signals. Experiments addressing the issues of the magnetic order, magon lifetime, and in general magnonics in two-dimensional AFMs would cast more light on the questions raised in this article.

\acknowledgements
Both authors thank Carl Trygger Stiftelse and Stiftelsen Olle Engkvist Byggm\"astare and J.F. thanks Vetenskapsr\aa det for funding. M. Sh. acknowlegdes Dr. Manuel Pereiri and  Dr. Fariborz Parhizgar for useful discussion.

\appendix
\section{Bogoliubov transformation}\label{AppA}
The eigen-modes of Hamiltonian Eq. \eqref{nonH} can be obtained by diagonalization through the following $2\times2$ unitary Bogoliubov transformation \cite{kittel1963quantum, nolting2009quantum, colpa1978diagonalization} for bosonic quasiparticles where,
\begin{subequations}
\begin{align}\label{BogoloMatrix}
 \begin{pmatrix}
    \alpha_{\bf q}
    \\
    \beta^\dag_{\bf q}
    \end{pmatrix}
    =&
    \begin{pmatrix}
    u_{\bf q} & -v_{\bf q}
    \\
    -v^*_{\bf q} & u^*_{\bf q}
    \end{pmatrix}
    \begin{pmatrix}
    a_{\bf q}
    \\
    b^\dag_{-\bf q}
    \end{pmatrix},
    \\ &
    u^2_{\bf q} = \frac{\varepsilon_0 + \varepsilon^\alpha_{\bf q}}{2 \varepsilon^\alpha_{\bf q}},
    \\&
    v^2_{\bf q} = \frac{\varepsilon_0 - \varepsilon^\beta_{\bf q}}{2 \varepsilon^\beta_{\bf q}}.
\end{align}
\end{subequations}
and they obey $|u_{\bf q}|^2 - |v_{\bf q}|^2 = 1$.
By this transformation, we rewrite the Hamiltonian Eq. \eqref{Hdiag} in terms of eignestates $\phi^\dag_{\bf q} = \begin{pmatrix}
    \alpha^\dag_{\bf q}
    &&
    \beta^\dag_{\bf q}
    \end{pmatrix}$ as,
\begin{align}
    \mathcal{H}^{(2)}
    =
    \sum_{\bf q}
    \begin{pmatrix}
    \alpha^\dag_{\bf q}
    &&
    \beta^\dag_{\bf q}
    \end{pmatrix}
    \begin{pmatrix}
    \varepsilon^\alpha_{\bf q} & 0
    \\
    0 & \varepsilon^\beta_{\bf q}
    \end{pmatrix}
    \begin{pmatrix}
    \alpha_{\bf q}
    \\
    \beta_{\bf q}
    \end{pmatrix},
\end{align}
up to the quadratic terms. The Green's function in terms of new spinor reads as $G = -i \langle\langle \phi_{\bf q} \phi^\dag_{\bf q} \rangle\rangle $ which by using the same procedure of equation of motion, we get
\begin{equation}\label{AFMgreenDiag}
\begin{split}
{\bf G}_0(\varepsilon,{\bf q})=&
\frac{1}{(\varepsilon-\varepsilon^\alpha_{\bf q})(\varepsilon-\varepsilon^\beta_{\bf q})}
\begin{pmatrix}
\varepsilon - \varepsilon^\beta_{\bf q} & 0
\\
0 & \varepsilon - \varepsilon^\alpha_{\bf q}
\end{pmatrix},
\end{split}
\end{equation}
where the magnon density of states is obtained,
\begin{equation}
n(\varepsilon,{\bf q})
=
\frac{-1}{\pi} \mathbf{Im} \, \Big[
\frac{1}{\varepsilon-\varepsilon^\alpha_{\bf q}}
+
\frac{1}{\varepsilon-\varepsilon^\beta_{\bf q}}
\Big],
\end{equation}
in which $\varepsilon \rightarrow \varepsilon + i 0^+$.
From this representation, we also obtain the same behaviour of magnon DOS which comes from this fact that observation does not depend on framework.

\section{On-site Green's function}\label{AppB}

Here, one can find the analytic calculation of on-site Green's function for sublattice $\mathcal{A}$
\begin{align}
g_0^{\mathcal{A}\mathcal{A}}
= &
\frac{1}{\Omega}
\int \frac{d^2\mathbf{q}}{(2 \pi)^2}
\frac{\varepsilon + \varepsilon_0 - \varepsilon_h }{ (\varepsilon - \varepsilon_h)^2 - \varepsilon_0^2 + z^2 J^2S^2 |\gamma_\mathbf{q}|^2 }
\\&
\approx \nonumber
\frac{1}{2 \, \pi \, \Omega}
\int_0^{q_c} q dq
\frac{\varepsilon + \varepsilon_0 - \varepsilon_h }{ (\varepsilon+ i 0^+ - \varepsilon_h )^2 - 4 z J \mathcal{K}_z S^2  - \zeta^2 q^2/ 2}
\\& \nonumber = 
- \frac{\varepsilon-\varepsilon_h + \varepsilon_0}{W^2}
\Bigg[ 
\ln{\bigg|1- \frac{\zeta^2 q^2_c/2}{(\varepsilon-\varepsilon_h)^2 -\tilde{\cal K}}\bigg|}
\\&\nonumber
+
i \, \pi \, 
\text{sgn} (\varepsilon-\varepsilon_h) \,
\Theta\Big((\varepsilon - \varepsilon_h)^2 - \tilde{\cal K}\Big) \,
\Theta\Big(\zeta^2 q^2_c/2-(\varepsilon-\varepsilon_h)^2+\tilde{\cal K}\Big)
\Bigg],
\end{align}
where to solve such integral $\int_{-\infty}^{\infty} dx \frac{1}{x+i \eta} f(x)$ we use Cauchy principal value as $\frac{1}{x + i \eta} = \mathcal{P} \frac{1}{x} - i \pi \delta(x)$.

\section{Ferromagnet}\label{AppC}
The effective model for bosonic excitations of spins interacting ferromagnetically on a honeycomb lattice can be read as \cite{PhysRevB.94.075401},
\begin{align}
\mathcal{H}_{FM}
=
\sum_{\bf q} \Big[ \varepsilon_0 a^\dag_{\bf q} a_{\bf q} + \varepsilon_0 b^\dag_{\bf q} b_{\bf q} 
- S Z J \gamma_{\mathbf{q}}  a^\dag_{\bf q} b_{\bf q} 
- S Z J \gamma_{-\mathbf{q}}  b^\dag_{\bf q} a_{\bf q} \Big],
\end{align}

\section{Square lattice}\label{AppD}

\begin{figure}[H]
\begin{center}
    {\includegraphics[width=0.49\columnwidth]{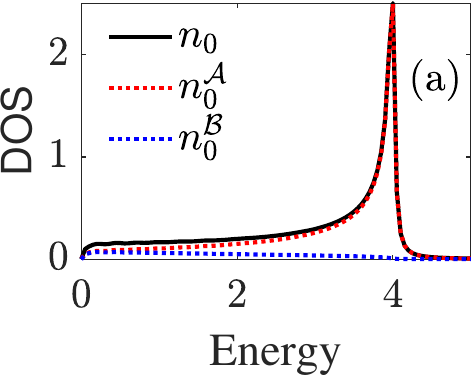}} 
    {\includegraphics[width=0.49\columnwidth]{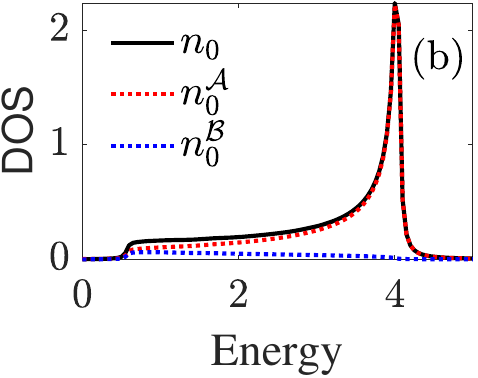}}
    {\includegraphics[width=0.49\columnwidth]{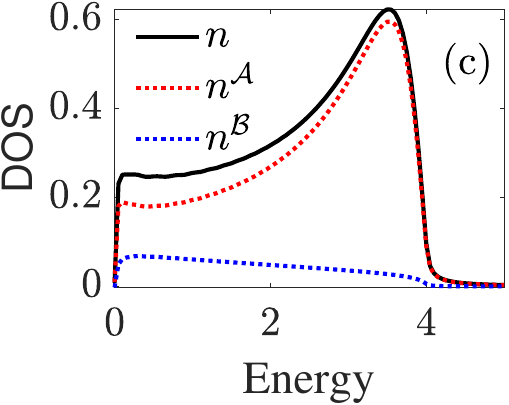}}
    {\includegraphics[width=0.49\columnwidth]{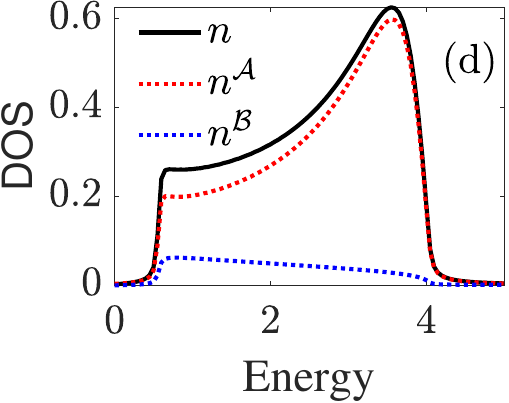}}
\end{center}
\caption{Integrated magnon DOS for the pristine and the perturbed AFM with a single defect on a square lattice. Panel (a,c) and (b,d) display magnon DOS for the isotropic and uniaxial AFMs with $K_z = 0.02J$. In these panels, solid black lines show the total DOS while red and blue dotted lines represent the contribution of sublattice $\mathcal{A}$ and $\mathcal{B}$ in total DOS. We use $J = 1$, $S=1$, $U_0 = -0.5$.}
\label{fig:afmsquare}
\end{figure}

In Fig. \eqref{fig:afmsquare}, the integrated magnon DOS for isotropic (a,b) and easy-axis (c,d) AFM on a square lattice are plotted. Panels (a,c) show the pristine case while panels (b,d) refer to the lattice including a scattering potential. We observe that the behaviour of magnon DOS in a square lattice is quite as same as the one in honeycomb lattice in both a pristine and perturbed system.

\bibliography{ref}
\end{document}